\documentclass[conference]{IEEEtran}
\IEEEoverridecommandlockouts
\usepackage{cite}
\usepackage{amsmath,amssymb,amsfonts}
\usepackage{algorithmic}
\usepackage{graphicx}
\usepackage{textcomp}
\usepackage[utf8]{inputenc}
\usepackage{xcolor}
\usepackage{url}
\usepackage{tikz}
\usepackage{booktabs}
\usepackage{placeins}
\def\BibTeX{{\rm B\kern-.05em{\sc i\kern-.025em b}\kern-.08em
    T\kern-.1667em\lower.7eCorex\hbox{E}\kern-.125emX}}
\begin{document}

\title{\textsc{Splice-Core}: Low-Order-Bit-Splicing GEMM for Highly Energy-Efficient AI Inference Accelerators}

\author{ \IEEEauthorblockN{ Prabhu Vellaisamy\IEEEauthorrefmark{1}, Hsien-Fu Hung\IEEEauthorrefmark{2}, Andrew Kim\IEEEauthorrefmark{1}, Shawn Blanton\IEEEauthorrefmark{1}, John Paul Shen\IEEEauthorrefmark{1}} \IEEEauthorblockA{\IEEEauthorrefmark{1} Carnegie Mellon University} \IEEEauthorblockA{\IEEEauthorrefmark{2} Massachusetts Institute of Technology}\IEEEauthorblockA{ \{pvellais, akim2, rblanton, jpshen\}@andrew.cmu.edu } \IEEEauthorblockA{ whung@mit.edu } }

\maketitle

\begin{abstract}
\textsc{Splice-Core} is a temporal-unary/binary 
(tub) hybrid general matrix multiplication (GEMM) unit that minimizes low-order accumulation latency, while retaining value-dependent execution for high-order activation bits. We present two novel processing element (PE) variants: (i) Splice-BP utilizing bit planes, and (ii) Splice-LUT, which selects a shared weight multiple in a single cycle. Compared to a binary systolic array, Splice-BP/Splice-LUT achieve 77.4\%/76.4\% geometric-mean tile energy reductions, respectively, across 26 INT4 convolutional neural network (CNN) and large language model (LLM) workloads in 45nm. Splice-LUT further increases geometric-mean tile throughput by 44.4\% and reduces geometric-mean energy-delay product (EDP) by 6.13$\times$ across the 26 configurations. Across the evaluated W8A8 convolution and expert-projection workloads, Splice-BP and Splice-LUT reduce tile energy by 77.3–88.1\%. Post-place-and-route INT4 yields 14\%/9\% smaller area for Splice-BP/Splice-LUT, compared to the systolic array.
\end{abstract}


\section{Introduction}

General matrix multiplication (GEMM) is a fundamental operation in convolutional networks and transformer feed-forward networks (FFNs). Quantization reduces operand storage and arithmetic cost by representing values with fewer bits~\cite{pact2018,smoothquant2023,quarot2024}, while pruning increases weight sparsity by setting selected weights to zero~\cite{deepcompression2015,wanda2023}. Temporal-unary designs use conditional accumulation to skip updates for zero operands and perform fewer updates for smaller encoded magnitudes~\cite{tugemm2023,tubgemm2023,tempus2025}. When processing element (PE) rows advance together, the largest encoded activation magnitude across the rows sets the duration of the temporal step. Rows with smaller magnitudes can stop accumulating earlier. This dependence on operand values helps explain why the efficiency of temporal-unary GEMMs varies with precision, array size, and operand distribution~\cite{unaryexploration2024}.

AI accelerators run GEMM using specialized matrix multiplication hardware. Google's Tensor Processing Units (TPUs) execute GEMM on systolic arrays within their matrix multiply units (MXUs), while NVIDIA GPUs use Tensor Cores. In LLM inference, these engines execute GEMMs in both the attention and feed-forward (FFN) networks. Attention-FFN disaggregation (AFD) places the two computations on separate worker groups so that each can be provisioned independently~\cite{megascale2025,stepfun2025step3,wu2026disaggregation}. On the FFN side, mixture-of-experts (MoE) layers increase model capacity while limiting per-token computation. A router assigns each token to a small subset of expert FFNs~\cite{deepseekv3}. This distributes the input batch across experts, producing smaller GEMMs of varying sizes. At low batch sizes, individual experts may receive too few tokens to fully utilize the matrix engine~\cite{megascale2025,nvidiagemm,taxbreak2026,vellaisamy2025characterizing}.

This highlights the need for GEMM engines that minimize arithmetic operations for padding and zero-valued operands (sparsity), while also reducing latency. We introduce \textsc{Splice-Core}, a temporal-unary/binary (tub) hybrid design for exact GEMM computation, featuring two PE variations. \textsc{Splice-Core} extends tubGEMM\cite{tubgemm2023} by replacing
temporal counting of low-order bits with a fixed-cycle update while retaining temporal-unary encoding for higher-order bits, with both phases sharing the common accumulator. Splice-BP (bitplane) updates one low-order bit plane during each cycle, whereas Splice-LUT selects a shared weight multiple for columns and completes the low-order update within a single cycle. Both variants inherently gate the accumulator update when the inputs are zero. Key contributions of this work include:
\begin{itemize}
\item \textbf{Exact temporal-unary/binary GEMM with improved speedup.} \textsc{Splice-Core} directly accumulates the low-order activation contribution and retains value-dependent temporal execution for the high-order bits. At same INT4 split, Splice-BP accumulates two low-order bit planes in two cycles, while Splice-LUT selects a column-shared weight multiple and accumulates it in one cycle (Section~\ref{sec:microarchitecture} and Fig.\ref{fig:splice-datapaths}). Both \textsc{Splice-Core} variants perform exact GEMM (not approximation), preserving accuracy. \textsc{Splice-Core} reduces accumulator switching on padded and sparse tiles by inherently gating inactive rows.
\item \textbf{INT4 synthesis evaluation across 26 workloads.} Across CNN, dense FFN, and routed-expert tiles, Splice-BP and Splice-LUT reduce geometric-mean energy by 77.4\% and 76.4\%, respectively, vs. the systolic array. Splice-LUT increases geometric-mean throughput by 44.4\% and improves the energy-delay product for all 26 configurations.
\item \textbf{Lower energy and higher throughput for INT8 vs. binary systolic array.} A prior study~\cite{unaryexploration2024} found that temporal-unary designs consumed more energy than binary GEMM at INT8 under worst-case latency. Splice-LUT reduces energy by 77.3\% for ResNet-50 convolutions and 88.1\% for OLMoE expert projections, and increases throughput by 22.6\% and 1.5\%, respectively, compared to the INT8 binary systolic array.
\item \textbf{Place-and-route and frequency scaling analysis.}
For INT4 place-and-route evaluation, the results show  14.4\%/9.0\% smaller areas for Splice-BP/Splice-LUT, respectively, compared to systolic array. In a separate 1\,GHz synthesis, Splice-BP and Splice-LUT consume 72.1\% and 72.7\% lower energy than the systolic array, respectively.
\end{itemize}

\section{Background}

\subsection{AI Inference Workloads and GEMM Dataflows}
Weight reuse is central to GEMM in both CNNs and transformers. Convolutional layers reuse filters across spatial positions~\cite{eyeriss2016}, and transformer FFNs reuse weight matrices across tokens by batching them into GEMMs. During decoding, the number of requests that can run concurrently depends partly on the capacity and bandwidth available for the request-specific key-value (KV) cache~\cite{distserve2024,pagedattention2023}. In MoEs, the router distributes tokens among expert FFNs~\cite{switch2022,deepseekv3}. Each expert therefore sees a potentially smaller batch, and its token count sets the row dimension of its GEMMs.

These different memory access patterns motivate attention-FFN disaggregation (AFD). Attention workers maintain the KV cache and exchange token activations with FFN workers~\cite{megascale2025,wu2026disaggregation}. An SRAM-rich FFN worker can stream activations through its compute engine while keeping weights on chip across batches. The Cerebras Wafer-Scale Engine and Groq LPUs illustrate how inference weights can reside in on-chip SRAM distributed across cores or devices~\cite{waferllm2025,ship2026}. Once a tile reaches the compute engine, the dataflow determines how weights and partial sums are reused within the array. In systolic engines, such as SRAM-rich devices or TPUs, data pass between neighboring processing elements. A weight-stationary dataflow holds a weight tile in the array as activations stream through, while an output-stationary dataflow keeps partial sums local until the reduction completes~\cite{tpuarchitecture,dip2025}. Buffering and pipelining help keep the array busy by overlapping operand transfers with computation and amortizing fill and drain overhead across tiles~\cite{pallaspipeline2026,cutlassgemm2026}. Array utilization depends on how much of each tile the batch fills. When an expert receives too few tokens to fill a tile, clock gating can reduce switching in the padded rows. \textsc{Splice-Core} suppresses accumulator operations for zero-valued and zero-padded activations, resulting in lower power and energy.

\subsection{Low-Precision Quantization - Temporal-Unary Arithmetic}
Low-precision quantization reduces the number of bits used to represent weights and activations in neural networks. The terms W4A4 and W8A8 refer to four-bit and eight-bit representations of weights and activations, respectively. 
SmoothQuant~\cite{smoothquant2023} rescales activations and weights to reduce the impact of outliers when inferring in W8A8 LLMs. Meanwhile, QuaRot~\cite{quarot2024} employs function-preserving rotations to facilitate W4A4 inference in LLMs.

Lower-precision bounds the cycle count of temporal-unary arithmetic, which encodes integer magnitudes as active cycle counts.
tuGEMM~\cite{tugemm2023} explores both exact serial and parallel temporal datapaths. tubGEMM~\cite{tubgemm2023} maintains one operand in binary form and conditionally accumulates it over the active duration of the other operand (temporal-unary/binary hybrid). Tempus Core~\cite{tempus2025} implements the same hybrid arithmetic approach for convolution operations in NVDLA. Although PE rows with smaller magnitudes may execute faster, a synchronized group progresses only after the row with the largest temporal magnitude has completed.

\subsection{Other Prior GEMM Designs}
Cambricon-C~\cite{cambricon2024} counts occurrences of 4-bit operand pairs and weights the counts by their corresponding products to compute dot products, avoiding repeated multiplications. uGEMM~\cite{ugemm2020} processes rate-coded and temporal-coded unary streams and supports early termination to trade numerical accuracy for lower latency and energy. DiP employs a systolic array with diagonally fed activations and permuted stationary weights, eliminating input and output synchronization FIFOs~\cite{dip2025}. ADiP uses reconfigurable PEs and shared accumulators to process multiple lower precision weight matrices in parallel with a shared activation matrix~\cite{adip2026}. OpenGeMM integrates a configurable GEMM core with a lightweight RISC-V controller and a banked scratchpad, using input prefetching and output buffering to improve array utilization~\cite{opengemm2025}. Comparative analysis is in Table~\ref{tab:prior-results}.

\section{\textsc{Splice-Core} Microarchitecture}\label{sec:microarchitecture}

\subsection{Array Organization and Operand Decomposition}
\begin{figure}[t]
    \centering
    \includegraphics[width=1.03\columnwidth]{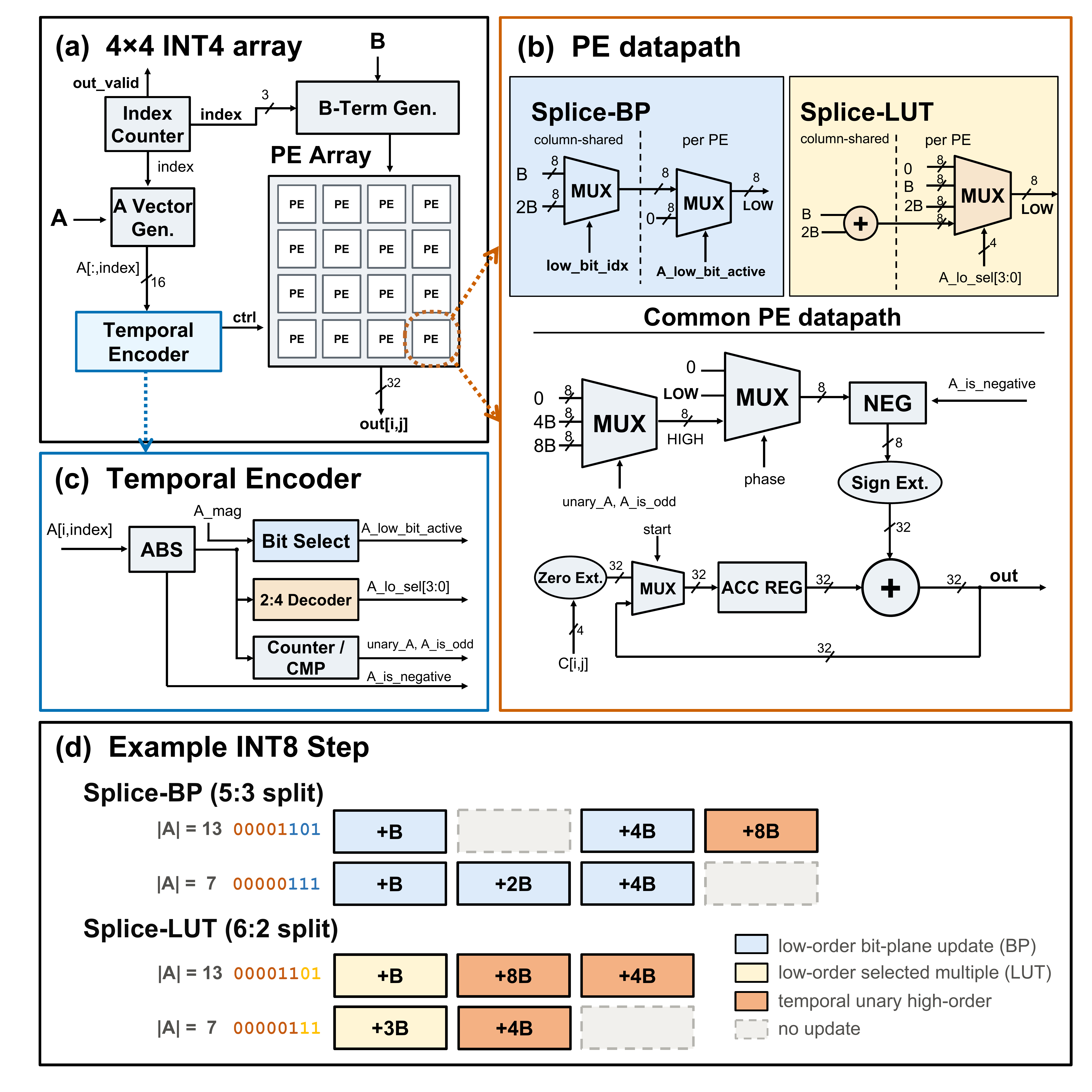}
    \caption{\textbf{Splice-Core microarchitecture.}
    (a) PE array with per-column weight-multiple generation and per-row broadcast of temporal control signals.
    (b) Splice-BP and Splice-LUT low-order datapaths and their common high-order accumulation datapath.
    (c) Temporal encoder generating low-order selection and high-order unary control signals. (d) Illustrating lower-order-bit-splicing for Splice-BP and Splice-LUT.}
    \label{fig:splice-datapaths}
\end{figure}

\textsc{Splice-Core} computes the matrix product $C =AB$ where the activation matrix $A \in \mathbb{R}^{M \times K}$ and the weight matrix $B \in \mathbb{R}^{K \times N}$. The resulting output matrix is $C \in \mathbb{R}^{M \times N}$. An $M_t\times N_t$ array of PEs computes one output tile, where $t$ denotes a tile dimension. At reduction step $k$, row $i$ broadcasts control signals derived from activation $A_{ik}$, and column $j$ broadcasts multiples of weight $B_{kj}$. PE $(i,j)$ uses the row signals to determine when to update and which weight term to add or subtract. Together, these updates accumulate the product $A_{ik}B_{kj}$ into the PE's local INT32 partial sum:
\begin{equation}
C_{ij}\leftarrow C_{ij}+A_{ik}B_{kj}
\end{equation}
Each reduction tile contributes $K_t$ outer products. The PEs retain their partial sums across reduction tiles until all $K$ terms have been accumulated. \textsc{Splice-Core} splits each activation magnitude at bit $\ell$ as 
$|A_{ik}|=2^\ell a_{\mathrm{hi}}+a_{\mathrm{lo}}$,
where $a_{\mathrm{lo}}$ contains the \(\ell\) least significant bits. With \(s=\operatorname{sgn}(A_{ik})\), the product is
\begin{equation}
A_{ik}B_{kj}
=s\left(2^\ell a_{\mathrm{hi}}B_{kj}
+a_{\mathrm{lo}}B_{kj}\right).
\label{eq:splice-update}
\end{equation}
The \textsc{Splice-Core} PEs accumulate the low-order bits in a fixed number of cycles, then compute the high-order contribution by temporally accumulating the weight left-shifted by $\ell$ bits. Each row stops updating once it completes its contribution, and the largest $a_{\mathrm{hi}}$ among the synchronized rows determines latency of a single GEMM tile. The binary weight $B_{kj}$ remains fixed throughout step $k$. A single unit in each column generates the weight multiples and shares them across all $M_t$ PEs in that column. Each PE uses selects its contribution, eliminating the need for a general-purpose multiplier in each PE. Both phases share the signed adder and INT32 accumulator shown in Fig.~\ref{fig:splice-datapaths}. This weight sharing occurs within an outer-product step, while the surrounding memory system handles buffering across steps.

During GEMM, a PE enables an update only when its activation control is asserted and its weight is nonzero. The update enables are ORed across each row to drive a shared accumulator clock gate. A latch captures the enable while the clock is low and holds it stable while the clock is high. This lets rows with no active contributions (including zero-padded rows) deassert accumulator clock pulses. Reset, initialization, and clearing also enable the gated clock.

\subsection{Splice-BP and Splice-LUT: Low-Order Accumulation}
Temporally processing the low-order activation bits requires repeated accumulation, with latency determined by the largest low-order magnitude across synchronized rows. The sparsity of quantized activations dictates that the high-order bits of these activations are often zero, while their low-order bits are more uniformly distributed. We developed two PE variants for \textsc{Splice-Core} that trade low-phase latency against hardware complexity. Splice-BP processes one low-order bit per cycle, writing the low-order magnitude as $a_{\mathrm{lo}}=\sum_{q=0}^{\ell-1}b_q2^q$. The PE conditionally accumulates the shifted weight $2^qB_{kj}$ when bit $b_q$ is asserted. Activation sign determines addition or subtraction. The shifted weights are generated once per column and shared across its PEs. Thus, the low phase takes exactly $\ell$ cycles, while zero bits do not make arithmetic updates. For the 2:2 INT4 split, this requires two cycles. Splice-BP also accommodates wider low-order slices without generating every possible weight multiple. Splice-LUT reduces the two-bit low phase to one cycle by selecting the complete low-order product $a_{\mathrm{lo}}B_{kj}\in
\{0,B_{kj},2B_{kj},3B_{kj}\}$.
Each column generates these multiples using a shift for $2B_{kj}$ and an addition for $3B_{kj}$. Each PE selects the required multiple using its row's low-order activation bits and adds or subtracts it from the accumulator. The Splice-LUT is implemented as combinational selection logic and requires no table memory or table-construction phase.
At the same two-bit INT4 split, Splice-BP uses two conditional updates, whereas
Splice-LUT uses one update at the cost of a wider selector and a shared $3B_{kj}$
generator. The separately synthesized variants retain the same high-order
schedule, so their INT4 comparison shows the cycle saving and implementation
cost of changing the low-order path. Weight multiples are widened before
shifting or addition to preserve exact signed arithmetic. For INT8, Splice-BP and Splice-LUT use a 5:3 and 6:2 split, respectively. Depending on the application and its activation distribution, either PE variant may be better positioned to balance the work between the low- and high-order phases. Whereas tubGEMM counts the full activation magnitude two units per cycle, requiring up to a worst case latency of $2^{w-2}$ per outer product, \textsc{Splice-core} counts only the high-order part after a fixed low-order phase of $L$ cycles
($L=\ell$ for Splice-BP and $L=1$ for Splice-LUT), bounding each computation at
$L + 2^{w-\ell-2}$ cycles. The one-cycle-per-outer-product reduction established for the common-split INT4 designs does not directly extend to INT8.

\section{Experimental Methodology}\label{sec:exp}

\textsc{Splice-Core} and a systolic array baseline are implemented in SystemVerilog and synthesized using Cadence Genus 21.19 with the Nangate45 library. We evaluate $16 \times 16$ and $32 \times 32$ arrays on $16 \times 16 \times 16$ and $32 \times 32 \times 32$ GEMM tiles, respectively. The operating frequency is set to 400 MHz, although we also assess performance at 1 GHz for frequency-scaling analysis. To provide an additional point of comparison, we include a row-gated systolic array that isolates one multiplier operand, preventing updates to the accumulator for all-zero rows. This setup allows the operand and pipelines to continue advancing, while update enables are utilized without necessitating an explicit clock gate.

Post-synthesis area and power reporting account for the signed INT32 accumulator datapaths within \textsc{Splice-Core} as well as the local systolic baselines. Power reports are derived from VCD activity across all evaluated tiles for each workload. In both scenarios, tile energy is determined by multiplying the mean power of the window by the mean tile duration for the configuration. The workload traces includes: (i) convolution tiles from MobileNetV2, MobileNetV3, and ResNet-50; (ii) Llama FFN prefill/decode; (iii) OPT ReLU projections; and (iv) expert projections from OLMoE and Qwen1.5-MoE. The Qwen traces include request batches of sizes 1, 4, 16, and 32. CNN inputs utilize resized CIFAR-10 test images, while transformer inputs draw from WikiText-2 validation contexts. Llama, OLMoE, and OPT operands leverage symmetric requantization based on observed peak magnitudes, whereas Qwen employs per-token activation quantization and group-128 weight quantization.

\section{Results and Analysis}\label{sec:results}
\subsection{INT4 Synthesis Analysis}

\begin{table}[!h]
\centering
\caption{INT4 post-synthesis area, power, and cycles at 400\,MHz.
Power is the geometric mean of per-case mean total power. Cycle ranges span the mean tile cycle counts across workloads.}
\label{tab:int4-implementation}
\footnotesize
\setlength{\tabcolsep}{3pt}
\begin{tabular}{@{}llrrr@{}}
\toprule
Tile & Design & \shortstack[r]{Cell area\\(mm$^2$)}
     & \shortstack[r]{Power\\(mW)} & Cycles/tile \\
\midrule
$16^3$ & Systolic   & 0.1254 &  32.46 & 48 \\
       & Splice-BP  & 0.1141 &   7.16 & 49--50.5 \\
       & Splice-LUT & 0.1183 &  10.87 & 33--34.5 \\
\addlinespace[2pt]
$32^3$ & Systolic   & 0.5152 & 128.45 & 96 \\
       & Splice-BP  & 0.4408 &  28.28 & 97--106.75 \\
       & Splice-LUT & 0.4711 &  44.22 & 65--74.75 \\
\bottomrule
\end{tabular}
\end{table}

\begin{figure}[!h]
\centering
\includegraphics[width=\columnwidth]{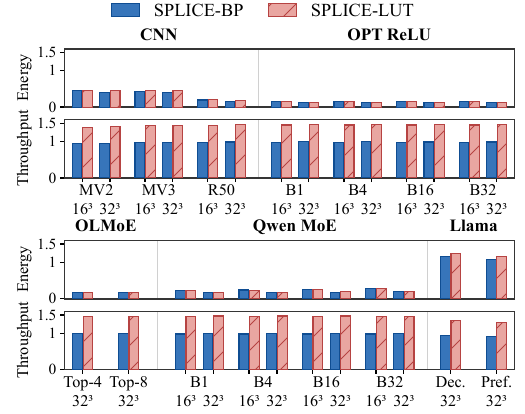}
\caption{INT4 tile energy and throughput at 400\,MHz across 26 workload configurations. Both metrics are normalized to the systolic array. MV2, MV3, and R50 denote MobileNetV2, MobileNetV3, and ResNet-50. B denotes request batch size.}
\label{fig:int4-workload-bars}
\end{figure}

\textbf{Tile energy and workload dependence.}
Fig.~\ref{fig:int4-workload-bars} shows geometric-mean energy reductions of 77.4\%/76.4\% for Splice-BP/Splice-LUT vs. systolic baseline, with savings in 24 of 26 workloads. The reductions are 68.3\%/65.8\% for CNNs, 86.4\%/86.1\% for OPT ReLU projections, and 79.8\%/79.3\% for routed-expert projections. Geometric-mean EDP improves by 4.31x and 6.13x for Splice-BP and Splice-LUT, respectively, across all 26 cases. For Llama workloads, Splice-BP/Splice-LUT use 16.5\%/24.5\% more tile energy on decode and 7.2\%/15.4\% more on prefill, relative to the systolic array. Compared with the heavily padded expert tiles, the Llama tiles have fewer zero activations, so \textsc{Splice-Core} gates its row accumulators less often, and larger high-order activation magnitudes extend the temporal phase. Further, the geometric mean of the row-gated systolic array's tile energy divided by that of \textsc{Splice-Core} is \(5.85\times\) for Splice-BP and \(5.70\times\) for Splice-LUT. The corresponding ratios for the ordinary systolic array are \(5.86\times\) and \(5.71\times\), respectively.


\textbf{Area and power.}
Table~\ref{tab:int4-implementation} summarizes 11 configurations with \(16^3\) tiles and 15 with \(32^3\) tiles. Compared to the systolic array, Splice-BP and Splice-LUT reduce cell area by 9.0\% and 5.7\%, respectively, for the \(16\times16\) PE array, and by 14.5\% and 8.6\% for the \(32\times32\) array. The geometric-mean reductions in total power are 78.0\%/66.0\% for Splice-BP/Splice-LUT. In the heavily padded Qwen B1 case with \(32^3\) tiles, dynamic power accounts for most of the savings. Splice-BP/Splice-LUT draw 12.08/23.45\,mW of dynamic power, respectively, compared with 120.57\,mW for systolic array. Leakage ranges from 9.48-9.80\,mW across the three designs.

\textbf{Cycle count and throughput.}
Splice-LUT replaces Splice-BP's two low-order updates with a single update, saving one cycle per outer product. This reduces tile latency by exactly 16 cycles for \(16^3\) tiles and 32 cycles for \(32^3\) tiles in every INT4 configuration. Splice-LUT also completes each of the 26 configurations in fewer cycles than the systolic array, increasing geometric-mean throughput by 44.4\%, while Splice-BP delivers 2.5\% lower throughput. Its smaller cell area yields 11\% higher throughput per cell area. Splice-LUT improves the same metric by 55.9\%.

\textbf{Splice-BP vs. Splice-LUT tradeoff.}
Across the 26 INT4 cases, Splice-LUT's single-cycle low-order update increases geometric-mean isolated-tile throughput by 48.2\% over Splice-BP. Splice-LUT draws 54.4\% more power on a geometric-mean basis, but its shorter latency limits the increase in tile energy to 4.2\% and reduces EDP by 29.7\%.

\subsection{INT8 Synthesis Results}

In Table~\ref{tab:int8}, we compare Splice-BP and Splice-LUT with the systolic baseline on ResNet-50 convolutions and OLMoE top-8 expert projections. Both workloads use W8A8 operands and \(32^3\) tiles, with switching activity captured over the full evaluation windows. Relative to systolic, Splice-BP/Splice-LUT reduce estimated tile energy by 77.8\%/77.3\% on ResNet-50, and by 86.3\%/88.1\% on OLMoE, respectively.

\begin{table}[!h]
\centering
\caption{INT8 post-synthesis results for $32^3$ tiles at 400\,MHz. Throughput counts two operations per MAC over padded tile.}
\label{tab:int8}
\footnotesize
\setlength{\tabcolsep}{3pt}
\begin{tabular}{@{}lrrrrr@{}}
\toprule
Design & Cell area & Power & Cycles & Energy & Throughput \tabularnewline
       & (mm$^2$)  & (mW) & per tile & (nJ) & (GOPS) \tabularnewline
\midrule
\multicolumn{6}{@{}l}{\textit{ResNet-50 convolution}} \tabularnewline
Splice-BP       & 0.5023 &  36.64 & 134.38 & 12.31 & 195.08 \tabularnewline
Splice-LUT      & 0.5260 &  64.20 &  78.33 & 12.57 & 334.65 \tabularnewline
Systolic & 0.7803 & 231.28 &  96.00 & 55.51 & 273.07 \tabularnewline
\addlinespace[2pt]
\multicolumn{6}{@{}l}{\textit{OLMoE top-8 expert projection}} \tabularnewline
Splice-BP       & 0.5023 &  26.87 & 140.75 &  9.46 & 186.25 \tabularnewline
Splice-LUT      & 0.5260 &  34.71 &  94.63 &  8.21 & 277.03 \tabularnewline
Systolic & 0.7803 & 287.76 &  96.00 & 69.06 & 273.07 \tabularnewline
\bottomrule
\end{tabular}
\end{table}

Splice-BP draws less power on both workloads, while Splice-LUT completes tiles in fewer cycles. On OLMoE, the shorter latency offsets Splice-LUT's higher power, yielding 13.2\% lower tile energy than Splice-BP. The latency difference reflects both the low-order slice width and the update mechanism, since the INT8 variants differ in both. Relative to the systolic array, Splice-LUT improves isolated-tile throughput by 22.6\% on ResNet-50 and 1.5\% on OLMoE. Splice-BP takes more cycles than the systolic array on both workloads. Further, Splice-BP and Splice-LUT occupy 0.5023\,mm\(^2\) and 0.5260\,mm\(^2\), respectively, vs. 0.7803\,mm\(^2\) for the systolic array. Splice-BP's smaller cell area offsets its lower throughput, yielding \(1.11\times\) and \(1.06\times\) the systolic throughput per cell area on ResNet-50 and OLMoE, respectively. Splice-LUT combines smaller cell area with higher throughput, reaching \(1.82\times\) and \(1.51\times\) the systolic throughput per cell area on the same workloads.

\begin{table*}[!t]
\caption{Comparison of GEMM implementations across precision,
technology, and evaluation scope.}
\label{tab:prior-results}
\centering
\scriptsize
\setlength{\tabcolsep}{2pt}
\renewcommand{\arraystretch}{1.05}
\begin{tabular*}{\textwidth}{@{\extracolsep{\fill}}lllllrlrrrr@{}}
\toprule
Design & Venue & Process & Precision & Geometry & Clock & Scope
  & Area & Power & Energy & Throughput \tabularnewline
 & & & & & (MHz) & & (mm$^2$) & (mW) & (nJ/tile)
 & (GOPS) \tabularnewline
\midrule
\multicolumn{11}{@{}l}{\textbf{4-bit results}} \tabularnewline
tubGEMM~\cite{unaryexploration2024} & ISVLSI 2024 & Nangate45 & 4-bit
  & $32\times32$ & 400 & Worst-case synth.
  & 0.1519 & 39.1 & 12.51 & 204.80$^{d}$ \tabularnewline
Cambricon-C~\cite{cambricon2024} & MICRO 2024 & FreePDK45 & W4A4
  & $32\times32$ & 400 & SA + converter
  & 12.23 & 1879.36 & -- & -- \tabularnewline
\addlinespace[2pt]
\multicolumn{11}{@{}l}{\textit{\textbf{This work: Qwen B1, $32^3$ tiles}}} \tabularnewline
Systolic & -- & Nangate45 & W4A4 & $32\times32$ & 400 & Array synth.
  & 0.5152 & 130.09 & 31.221 & 273.07 \tabularnewline
SPLICE-BP & -- & Nangate45 & W4A4 & $32\times32$ & 400 & Array synth.
  & \textbf{0.4408} & \textbf{21.55} & \textbf{5.239} & 269.62 \tabularnewline
SPLICE-LUT & -- & Nangate45 & W4A4 & $32\times32$ & 400 & Array synth.
  & 0.4711 & 33.24 & 5.421 & \textbf{401.90} \tabularnewline
\midrule
\multicolumn{11}{@{}l}{\textbf{8-bit results}} \tabularnewline
uGEMM~\cite{ugemm2020} & ISCA 2020 & TSMC 45\,nm & Bipolar 8-bit
  & $16\times16$ & 400 & Unscaled synth.
  & 0.770 & 200 & 130 & 12.80$^{a}$ \tabularnewline
tubGEMM~\cite{unaryexploration2024} & ISVLSI 2024 & Nangate45 & 8-bit
  & $32\times32$ & 400 & Worst-case synth.
  & 0.3387 & 90.9 & 465.41 & 12.80$^{d}$ \tabularnewline
DiP~\cite{dip2025} & IEEE TCAS-I & 22\,nm & W8A8
  & $32\times32$ & 1000 & Routed core
  & 0.253 & 211.5 & 13.54$^{c}$ & 1024$^{c}$ \tabularnewline
ADiP~\cite{adip2026} & IEEE OJSSCS & 22\,nm & W8A8
  & $64\times64$ & 1000 & Routed core
  & 1.32 & 1452$^{e}$ & -- & 8192$^{b}$ \tabularnewline
OpenGeMM~\cite{opengemm2025} & ASP-DAC 2025 & 16\,nm & W8A8
  & $8\times8\times8$ & 200 & System synth.
  & 0.531 & 43.8 & -- & 204.8$^{b}$ \tabularnewline
\addlinespace[2pt]
\multicolumn{11}{@{}l}{\textit{\textbf{This work: OLMoE top-8, $32^3$ tiles}}} \tabularnewline
Systolic & -- & Nangate45 & W8A8 & $32\times32$ & 400 & Array synth.
  & 0.7803 & 287.76 & 69.063 & 273.07 \tabularnewline
SPLICE-BP & -- & Nangate45 & W8A8 & $32\times32$ & 400 & Array synth.
  & \textbf{0.5023} & \textbf{26.87} & 9.456 & 186.25 \tabularnewline
SPLICE-LUT & -- & Nangate45 & W8A8 & $32\times32$ & 400 & Array synth.
  & 0.5260 & 34.71 & \textbf{8.212} & \textbf{277.03} \tabularnewline
\bottomrule
\end{tabular*}
\par\smallskip\raggedright\scriptsize
\textbf{Bold} indicates the best result within each matched local group.
$^{a}$Derived from reported tile latency.
$^{b}$Peak throughput.
$^{c}$Derived using DiP's 64-cycle analytical tile latency at 1\,GHz;
energy assumes reported core power over that interval.
$^{d}$Derived from tubGEMM's worst-case cycle model,
$32\cdot 2^{w-2}$ for operand width $w$.
$^{e}$Reported ADiP core power used for its peak-efficiency reference.
Local throughput counts two operations per MAC over an isolated padded
tile. OpenGeMM includes controller and scratchpad; local results cover
the compute array.
\end{table*}

\subsection{INT4 Frequency Scaling}

In addition to 400\,MHz synthesis, we also evaluate at 1\,GHz for INT4 trade-offs under a tighter timing constraint. We use Qwen MoE batch-1 decode with \(16^3\) tiles at both clock targets. Table~\ref{tab:freq} reports cell area, power, latency, and energy. At both targets, the mean cycle counts per tile are 49.078 for Splice-BP, 33.078 for Splice-LUT, and 48 for the systolic baseline. Because the cycle counts remain unchanged, increasing the clock from 400\,MHz to 1\,GHz increases the calculated isolated-tile throughput of each design by \(2.5\times\).

\begin{table}[!h]
\centering
\caption{INT4 synthesis at two clock targets for Qwen MoE batch-1
decode (\(16^3\) tiles). Slack is rounded to the nearest picosecond.}
\label{tab:freq}
\footnotesize
\setlength{\tabcolsep}{2.5pt}
\begin{tabular}{@{}llrrrrr@{}}
\toprule
MHz & Design & Area & Power & Latency & Energy & Slack \tabularnewline
&        & (mm\(^2\)) & (mW) & (ns) & (nJ) & (ps) \tabularnewline
\midrule
400 & Systolic & 0.12540 & 32.966 & 120.000 & 3.956 & 8 \tabularnewline
& Splice-BP       & 0.11412 &  7.308 & 122.695 & 0.897 & 0 \tabularnewline
& Splice-LUT      & 0.11831 & 10.746 &  82.695 & 0.889 & 0 \tabularnewline
\addlinespace[2pt]
1000 & Systolic   & 0.14211 & 79.716 & 48.000 & 3.826 & 1 \tabularnewline
& Splice-BP       & 0.14260 & 21.783 & 49.078 & 1.069 & 0 \tabularnewline
& Splice-LUT      & 0.15556 & 31.627 & 33.078 & 1.046 & 0 \tabularnewline
\bottomrule
\end{tabular}
\end{table}

\textbf{Area and energy.}
Increasing the clock target from 400\,MHz to 1\,GHz increases cell area by 25\% for Splice-BP, 31.5\% for Splice-LUT, and 13.3\% for the systolic array. Power also increases for all three designs. For systolic, the shorter tile duration offsets the increase in power, reducing estimated tile energy by 3.3\%. In contrast, estimated tile energy increases by 19.2\% for Splice-BP and 17.7\% for Splice-LUT. Nevertheless, at 1\,GHz, Splice-BP and Splice-LUT consume 1.069\,nJ and 1.046\,nJ per tile, corresponding to 72.1\% and 72.7\% lower energy than systolic, respectively.

\textbf{Area-latency tradeoff.}
Splice-LUT's cell-area advantage at 400\,MHz becomes a 9.5\% area overhead at 1\,GHz, where it occupies 0.156\,mm\(^2\) compared with 0.142\,mm\(^2\) for the systolic array. Despite the area overhead, Splice-LUT completes a tile in 33.1\,ns vs. 48\,ns for the systolic array, delivering \(1.45\times\) the isolated-tile throughput. All three designs meet the reported 1\,GHz synthesis constraint.

\subsection{INT4 Place-and-Route Results}
Table~\ref{tab:pnr} reports place-and-route for the \(32\times32\) arrays at a 400\,MHz target. All three designs use the same Innovus P\&R script. The input netlists retain the synthesis settings described in Section~\ref{sec:exp}. Splice-BP and Splice-LUT occupy 14.4\% and 9.0\% less routed cell area than the systolic array, respectively. Setup slack is \(+23\)\,ps for Splice-LUT and rounds to \(0\)\,ps for Splice-BP. The systolic array violates the setup target by \(-5\)\,ps. For each workload, Table~\ref{tab:pnr} pairs the power values with post-synthesis tile energy and throughput.

\begin{table}[!h]
\centering
\caption{INT4 P\&R results and workload metrics for
$32^3$ tiles at 400\,MHz. $\Delta A$ is the cell-area change from
synthesis to routing. Area and worst setup and hold slack apply to
both workloads within each design.}
\label{tab:pnr}
\scriptsize
\setlength{\tabcolsep}{2pt}
\begin{tabular}{@{}llrrrrrrr@{}}
\toprule
Design & Workload & Routed area & $\Delta A$ & Setup & Hold
       & Power & Energy & Throughput \\
       & & (mm$^2$) & (\%) & (ps) & (ps) & (mW) & (nJ/tile) & (GOPS) \\
\midrule
Splice-BP & OPT B1        & 0.439211 & $-0.36$ & $0$   & $+33$ & 19.30 &  3.78 & 270.25 \\
   & Llama prefill &          &         &       &       & 18.65 & 35.49 & 245.57 \\
\addlinespace[2pt]
Splice-LUT & OPT B1        & 0.467178 & $-0.84$ & $+23$ & $+19$ & 27.64 &  3.84 & 403.30 \\
    & Llama prefill &          &         &       &       & 25.66 & 38.22 & 350.69 \\
\addlinespace[2pt]
Systolic & OPT B1        & 0.513112 & $-0.41$ & $\mathbf{-5}$ & $+6$ & 103.71 & 31.15 & 273.07 \\
         & Llama prefill &          &         &                &      & 103.49 & 33.12 & 273.07 \\
\bottomrule
\end{tabular}
\end{table}

\subsection{Comparison with Prior Work}
Table~\ref{tab:prior-results} summarizes published and local operating points at 4 and 8 bits. At 4 bits, Cambricon-C reports 12.23\,mm\(^2\) and 1879.36\,mW for a \(32\times32\) array at 400\,MHz in FreePDK45, including its count converter~\cite{cambricon2024}. Cambricon-C's counter-based computation requires a separate conversion stage, whose energy overhead is better amortized over larger matrix computations. In contrast, \textsc{Splice-Core} directly accumulates numerical contributions using magnitude-dependent execution, without requiring a separate output converter. At the same array size and clock, tubGEMM occupies 0.1519\,mm\(^2\) in Nangate45 and reports 12.51\,nJ per tile at worst-case latency~\cite{unaryexploration2024}. In our Qwen B1 evaluation, Splice-LUT occupies 0.4711\,mm\(^2\) and draws 33.24\,mW. On the same Qwen B1 operands, Splice-LUT reduces tile energy from 31.221 to 5.421\,nJ and increases isolated-tile throughput from 273.07 to 401.90\,GOPS vs. the systolic baseline.

At 8 bits, the 22\,nm DiP core reports 0.253\,mm\(^2\) and 211.5\,mW at 1\,GHz~\cite{dip2025}. The 16\,nm OpenGeMM system reports a peak throughput of 204.8\,GOPS at 200\,MHz and draws 43.8\,mW including the controller and scratchpad~\cite{opengemm2025}. In Nangate45 at 400\,MHz, Splice-BP occupies 0.5023\,mm\(^2\) and draws 26.87\,mW on OLMoE top-8. On the same workload, Splice-LUT reduces tile energy from 69.063 to 8.212\,nJ relative to our systolic baseline, with comparable isolated-tile throughput (277.03 vs. 273.07\,GOPS). The published 8-bit tubGEMM point reports 465.41\,nJ per tile at worst-case latency~\cite{unaryexploration2024}. The same study reports 7.06\,nJ at INT4 and 25.70,nJ at INT8 for its binary outer-product engine on \(32^3\) tiles at 400\,MHz~\cite{unaryexploration2024}. Clock frequency, workload, and evaluation scope vary across these published points. uGEMM evaluates unary-stream arithmetic~\cite{ugemm2020}, while tuGEMM and tubGEMM use binary output accumulation~\cite{tugemm2023,tubgemm2023}. A common signed INT32 accumulator cost for these prior designs remains unverified. \textit{Our area and power estimates explicitly include the signed INT32 datapath.}

\textbf{Energy across operating points.}
Table~\ref{tab:prior-results} shows that \textsc{Splice-Core} achieves the lowest listed tile energy at both precisions, even with older process nodes in some comparisons. In matched comparisons with the local systolic array at 400\,MHz in Nangate45, Splice-LUT reduces tile energy by 82.6\% on INT4 Qwen B1 and by 88.1\% on INT8 OLMoE top-8. On the OLMoE tiles, the INT8 Splice-LUT array achieves 7.98\,TOPS/W. Published peak-based efficiencies are 5.64\,TOPS/W for 22\,nm ADiP at W8A8, 9.55\,TOPS/W for the \(64\times64\) 22\,nm DiP array, and 4.68\,TOPS/W for the 16\,nm OpenGeMM system~\cite{adip2026,dip2025,opengemm2025}. These figures differ in clock frequency, workload activity, and the hardware included in the power estimate.

For each published operating point, we looked for absolute tile energy or for power and latency for the same tile. Cambricon-C reports workload-level efficiency and leaves absolute \(32^3\) tile energy and throughput unspecified for the listed array-plus-converter configuration~\cite{cambricon2024}. ADiP reports peak W8A8 efficiency and relative energy for attention workloads, leaving absolute tile energy unspecified for the listed core~\cite{adip2026}. OpenGeMM reports power for a \(32^3\) block and peak throughput, with the execution time of the power test unspecified~\cite{opengemm2025}. For DiP, we derive 13.54\,nJ per tile by multiplying the reported core power by the analytical tile latency, assuming that the reported power applies throughout that interval~\cite{dip2025}.



\subsection{Sustained Throughput per Area}
At a fixed clock frequency and tile size, sustained throughput per cell area scales as \(1/(A_{\mathrm{cell}}N_{\mathrm{cyc}})\), where \(A_{\mathrm{cell}}\) is the array's cell area and \(N_{\mathrm{cyc}}\) is the average number of cycles between tile starts. To match the systolic array's throughput per cell area, Splice-BP and Splice-LUT can take up to 9.9\% and 6.0\% more cycles between starts for \(16^3\) tiles, and 16.9\% and 9.4\% more for \(32^3\) tiles, respectively.

We use each \textsc{Splice-Core} variant's mean tile-cycle count as the modeled interval between tile starts. When the systolic array starts a tile every \(2K_t\) cycles, Splice-LUT achieves \(0.963\times\) the systolic throughput at both tile sizes. Its smaller area brings throughput per cell area to \(1.021\times\) and \(1.053\times\) the systolic value for \(16^3\) and \(32^3\) tiles, respectively. Splice-BP reaches \(0.715\times\) and \(0.759\times\), respectively. Both variants fall below parity in throughput per cell area when the systolic array starts a tile every $K_t$ cycles. Under the $2K_t$ schedule, Splice-LUT would use less energy per completed tile if its average power during back-to-back execution were below 96.3\% of the systolic array's average power.

\section{Conclusions}\label{sec:conclusions}
Bounding the low-order phase improves tile latency while retaining the ability to shorten high-order execution for small magnitudes. Since most quantized activations are zero or near zero, the higher-order phase is often short or skipped entirely, reducing tile energy by more than 76\% compared with a binary systolic array.
Llama crossover shows why workload-dependent power and latency should be assessed together when selecting between GEMM datapaths. Splice-BP generally favors INT4 array energy and area, while Splice-LUT favors isolated-tile latency. In an SRAM-rich implementation, shorter tile residency could also reduce buffer energy. 


\bibliographystyle{IEEEtran}
\bibliography{refs}

\end{document}